\PassOptionsToPackage{dvipsnames}{xcolor}
\PassOptionsToPackage{expansion, protrusion}{microtype}
\documentclass[sigconf]{acmart}

\copyrightyear{2023} 
\acmYear{2023} 
\setcopyright{usgovmixed}
\acmConference[ICS '23]{2023 International Conference on Supercomputing}{June 21--23, 2023}{Orlando, FL, USA}
\acmBooktitle{2023 International Conference on Supercomputing (ICS '23), June 21--23, 2023, Orlando, FL, USA}
\acmPrice{15.00}
\acmDOI{10.1145/3577193.3593706}
\acmISBN{979-8-4007-0056-9/23/06}

\usepackage{amsmath,amsfonts}
\usepackage{algorithmic}
\usepackage{graphicx}
\usepackage{textcomp}
\usepackage[ruled,vlined,linesnumbered]{algorithm2e}
\usepackage{subfiles}
\usepackage{bm}
\usepackage{pgf}
\usepackage{soul}
\usepackage[utf8]{inputenc}
\usepackage[T1]{fontenc}
\usepackage{hyperref}
\usepackage{url}
\usepackage{booktabs}
\usepackage{siunitx}        
\usepackage{amsfonts}
\usepackage{nicefrac}
\usepackage[]{microtype}      
\usepackage{listings}
\usepackage{enumitem}
\usepackage[noabbrev]{cleveref}
\usepackage{stfloats}
\usepackage{varwidth}
\usepackage{multicol}
\usepackage{multirow}
\usepackage{makecell}
\usepackage{float}
\usepackage{listings}
\usepackage{caption}
\usepackage{subcaption}
\usepackage{tikz}
\usepackage{ifthen}
\usepackage{graphicx, xcolor}               
\definecolor{BLUE}{rgb}{.0, .2, .6}     
\definecolor{BLUE2}{HTML}{1e50a2}       
\definecolor{RED}{HTML}{c9171e}         
\definecolor{RED2}{HTML}{D7003A}        
\definecolor{INK}{HTML}{595857}         
\definecolor{YELLOW}{HTML}{f1c40f}      
\definecolor{GREEN}{rgb}{0,0.6,0}       
\definecolor{MIDGRAY}{rgb}{0.5,0.5,0.5} 

\definecolor{B}    {HTML}{2b66d3}
\definecolor{B2}   {HTML}{003399}
\definecolor{R}    {HTML}{c9171e}
\definecolor{R2}   {HTML}{d7003a}
\definecolor{INK}  {HTML}{595857}
\definecolor{Y}    {HTML}{f1c40f}
\definecolor{G}    {HTML}{009a00}
\definecolor{GRAY} {HTML}{808080}
\definecolor{MAUVE}{HTML}{9400D1}

\newcommand*\Circled[1]{
    \tikz[baseline=(char.base)]{%
        \node[shape=circle, draw=none, fill=gray!40, thick, inner sep=0.6pt] (char) {%
            \textcolor{black}{\sffamily#1}}; }}

\usepackage{listings}						
\newcommand{\FIG}{Figure }
\newcommand{\TAB}{Table }

\hypersetup{colorlinks=true}

\usetikzlibrary{backgrounds,scopes}

\lstdefinestyle{cuda}{
    belowcaptionskip=1\baselineskip,
    breaklines=true,
    xleftmargin=\parindent,
    language=C,
    showstringspaces=false,
    basicstyle=\footnotesize\ttfamily,
    keywordstyle=\bfseries\color{PineGreen},
    commentstyle=\color{white!40!black},
    identifierstyle=\color{NavyBlue},
    stringstyle=\color{orange},
    escapeinside={(*@}{@*)},
    morekeywords={shared, is, to},
    numbers=left,
    morekeywords={[2]{copy,syncthreads,nnz}},
    keywordstyle={[2]{\color{Purple}}},
}
\def\BibTeX{{\rm B\kern-.05em{\sc i\kern-.025em b}\kern-.08em
T\kern-.1667em\lower.7ex\hbox{E}\kern-.125emX}}

\usepackage{tabu}

\newcommand{\thiswork}{\textsc{gpuLZ}}
\newcommand{\cusz}{{cuSZ}}
\newcommand{\culzss}{CULZSS}

\begin{document}
\title[\thiswork]{\thiswork: Optimizing LZSS Lossless Compression for Multi-byte Data on Modern GPUs}

\newcommand{\iu}{Indiana University}
\newcommand{\anl}{Argonne National Laboratory}
\newcommand{\microsoft}{Microsoft}

\newcommand{\AFFIL}[4]{%
	\affiliation{%
		\institution{\small #1}
		\city{#2}\state{#3}\country{#4}
	}
}

\colorlet{THEMECOLOR}{B}

\newcommand{\IU}{\AFFIL{\iu}{Bloomington}{IN}{USA}}
\newcommand{\ANL}{\AFFIL{\anl}{Lemont}{IL}{USA}}

\settopmatter{authorsperrow=4}

\author{Boyuan Zhang}{\IU}
\email{bozhan@iu.edu}

\author{Jiannan Tian}{\IU}
\email{jti1@iu.edu}

\author{Sheng Di}{\ANL}
\email{sdi1@anl.gov}

\author{Xiaodong Yu}{\ANL}
\email{xyu@anl.gov}

\author{Martin Swany}{\IU}
\email{swany@indiana.edu}

\author{Dingwen Tao}{\IU}
\authornote{Corresponding author: Dingwen Tao, Department of Intelligent Systems Engineering, Luddy School of Informatics, Computing, and Engineering, Indiana University.
}
\email{ditao@iu.edu}

\author{Franck Cappello}{\ANL}
\email{cappello@mcs.anl.gov}

\newcommand{\visualfactor}{1.8}
\newcommand{\notavailable}{\color{gray!80}{n/a}}
\newcommand{\TP}[2]{\ \ %
	\pgfmathsetmacro{\advantage}{((#1-#2)/#2)*\visualfactor}%
	\pgfmathsetmacro{\lensmall}{\visualfactor}%
	\pgfmathsetmacro{\ratioraw}{#1/#2}%
	\begin{tabular}{@{} r @{}}
		\color{THEMECOLOR}\num[round-mode=places,round-precision=1]{#1}\makebox[2em][r]{\textcolor{gray!80}{\num[round-mode=places,round-precision=1]{#2}}} \\[-1.4ex]
		{\color{black}\sffamily\tiny\num[round-mode=places,round-precision=1]{\ratioraw}$\times$}
		{\setlength{\fboxsep}{0pt}%
			{\color{THEMECOLOR}\rule{\advantage em}{.22em}}%
			{\color{gray!50}\rule{\lensmall em}{.22em}}}
	\end{tabular}%
}

\renewcommand{\shortauthors}{Zhang et al.}

\begin{abstract}
	Today's graphics processing unit (GPU) applications produce vast volumes of data, which are challenging to store and transfer efficiently. Thus, data compression is becoming a critical technique to mitigate the storage burden and communication cost. LZSS is the core algorithm in many widely used compressors, such as Deflate. However, existing  GPU-based LZSS compressors suffer from low throughput due to the sequential nature of the LZSS algorithm. Moreover, many GPU applications produce multi-byte data (e.g., int16/int32 index, floating-point numbers), while the current LZSS compression only takes single-byte data as input. To this end, in this work, we propose \thiswork{}, a highly efficient LZSS compression on modern GPUs for multi-byte data. The contribution of our work is fourfold: First, we perform an in-depth analysis of existing  LZ compressors for GPUs and investigate their main issues. Then, we propose two main algorithm-level optimizations. Specifically, we (1) change prefix sum from one pass to two passes and fuse multiple kernels to reduce data movement between shared memory and global memory, and (2) optimize existing pattern-matching approach for multi-byte symbols to reduce computation complexity and explore longer repeated patterns. Third, we perform architectural performance optimizations, such as maximizing shared memory utilization by adapting data partitions to different GPU architectures. Finally, we evaluate \thiswork{} on six datasets of various types with NVIDIA A100 and A4000 GPUs. Results show that \thiswork{} achieves up to 272.1$\times$ speedup on A4000 and up to 1.4$\times$ higher compression ratio compared to state-of-the-art solutions.
\end{abstract}

\ccsdesc[500]{Theory of computation~Massively parallel algorithms}
\ccsdesc[500]{Theory of computation~Data compression}

\keywords{Lossless compression; LZSS; GPU; performance.}

\maketitle

\section{Introduction}\label{sec:intro}

Many applications running on high-performance parallel and distributed systems generate large amounts of data, which  leads to storage bottlenecks due to limited capacity.
Meanwhile, interconnect technologies in distributed systems advance relatively more slowly than computing power, causing inter-node communication and I/O bottlenecks to become a severe issue~\cite{use-case-Franck}.
This motivates the design of software solutions to increase the interconnect bandwidth, such as communication-avoiding linear algebra~\cite{anderson2011communication, georganas2012communication}.

Data compression is a popular solution to reduce communication and I/O overheads significantly. For example, due to the high data reduction capabilities, lossy compression has recently been extensively studied to alleviate I/O bottlenecks in large-scale distributed applications such as high-performance computing (HPC) simulations.
Since the saved data is often used for post-analysis and visualization, errors introduced by error-bounded lossy compression are acceptable for many applications~\cite{jin2020understanding, jin2021adaptive, dmitriev2022error, poppick2020statistical, wang2022tac, use-case-Franck}.

However, lossy compression may not be applicable for inter-node communication in most distributed applications since data is usually exchanged between nodes at least once per time step, resulting in an accumulation of compression errors beyond the acceptable level. This is especially important for HPC simulations where numerical stability is critical, as accumulated compression errors can affect the correctness of the results.

Unlike lossy compression, lossless compression can avoid the loss of accuracy despite the relatively low compression ratio. In practice, among many lossless compression algorithms, LZ-series lossless compression is one of the most important algorithms.
It can identify repeated subsequences/patterns, thereby reducing spatial redundancy of the input sequence.
Specifically, LZSS~\cite{storer1982data} is a derivative of the classical LZ77 algorithm~\cite{ziv1977universal} (i.e., the first LZ compression algorithm). It holds a sliding window for the input stream to search for the longest match and then encodes each match as one pointer, including its length and offset (will be discussed in \S\ref{sec:background}). Input data with longer repeated subsequences are more likely to achieve higher compression ratios with LZSS.
As an entropy coder, LZSS is often combined with other types of lossless coders to (e.g., with Huffman encoding as Deflate~\cite{deutsch1996rfc1951}) remove both spatial and frequency redundancy.

On one hand, multi-byte data such as long integers and floating-point numbers are common as input to lossless compression~\cite{takeda2002processing, wong2012full,najam2014multi}. However, the classic LZSS compression only takes a single byte as the input unit, ignoring the data characteristics of different data types. Using multiple bytes as units in LZSS can improve both compression throughput (due to fewer symbols to process) and ratio (due to longer repeated patterns).

On the other hand, more and more applications are being implemented on the GPU due to its high performance and energy efficiency~\cite{evans2022survey}, resulting in multiple critical use cases of GPU compression.
For example, GPU compression can speed up GPU-CPU data transfers~\cite{weissenberger2021accelerating}.
It can also reduce GPU memory footprint to support larger input in deep learning~\cite{jin2021comet}.
However, it is challenging to parallelize LZSS on GPUs due to its strong data dependency~\cite{storer1982data}. Simply chunking data and distributing them to different GPU threads would cause warp divergence~\cite{xiang2014warp}.

{\culzss}~\cite{ozsoy2011culzss} is a state-of-the-art open-source GPU LZSS compressor. It can achieve relatively higher compression throughput on the GPU than the CPU solution~\cite{fastlz}. However, {\culzss} faces several critical issues:
\Circled{1} It cannot handle multi-byte data, and simply modifying its algorithm to accommodate multi-byte input may result in a significant drop in compression ratio.
\Circled{2} It lacks tuning of parameters such as block size and sliding window size for different GPU architectures.
\Circled{3} Its encoding process is performed on the CPU, which introduces high CPU-GPU data movement overhead.

To solve the above issues, we propose a highly optimized LZSS compression for multi-byte data on modern GPUs (called \thiswork\footnote{The code is available at \url{https://github.com/hipdac-lab/GPULZ}.}). Specifically,
we deeply analyze {\culzss} and identify its performance issues.
Based on these issues, we propose two main algorithm-level optimizations and a series of performance optimizations.
These optimizations can improve compression throughput and ratio simultaneously.
To the best of our knowledge, \textit{this is the first work that optimizes LZSS compression for multi-byte data on GPUs}.

The main contributions of this paper are summarized as follows.

\begin{itemize}[noitemsep, topsep=2pt, leftmargin=1.3em]
	\item We develop a highly efficient LZSS compression on GPUs for multi-byte data. We perform an in-depth analysis of {\culzss} and investigate its main performance issues.
	\item We optimize the prefix sum from one pass to two passes and fuse multiple kernels (e.g., matching and local prefix sum) to reduce data movement between shared memory and global memory.
	\item We propose a pattern-matching method for multi-byte data, which can reduce computational complexity and explore longer repeated patterns.
	\item We propose a data partitioning method that can adapt to different GPU architectures to maximize shared memory utilization.
	\item We evaluate \thiswork{} on six datasets with NVIDIA A100 and A4000 GPUs. The evaluation demonstrates that {\thiswork} outperforms {\culzss} by up to 272.1$\times$ in compression throughput with no degradation of compression ratio (even 20.6\% improvement).
\end{itemize}

In \S\ref{sec:background}, we present the background about CUDA architecture, LZSS algorithm, GPU implementations of LZSS, and their issues. In \S\ref{sec:design}, we present the design of \thiswork{} with our algorithm-level and architectural performance optimizations. In \S\ref{sec:evaluation}, we evaluate \thiswork{} and compare it with other GPU LZ compression. In \S\ref{sec:conclusion}, we conclude the paper and discuss our future work.

\section{Background and Motivation}\label{sec:background}

In this section, we present the background of CUDA architecture, LZSS algorithm, and its state-of-the-art GPU implementations.

\subsection{CUDA Architecture}
CUDA is a parallel computing platform and API that allows the software to use NVIDIA GPUs for general-purpose processing.
Thread is the basic programmable unit for GPU programmers to use massive numbers of CUDA cores. CUDA threads are organized into three levels, grid, block, and thread.
Specifically, a group of 32 threads is called a \textit{warp}. All threads in the same warp will execute the same instruction. However, if different threads in a warp follow different control paths, some threads are masked from performing any useful work. This situation is called \textit{warp divergence}, which is one of the fundamental factors that limit the performance of GPUs. Multiple warps are combined to form a thread \textit{block}, and a set of thread blocks form a thread \textit{grid}.

Regarding the CUDA memory hierarchy, the largest and slowest memory is called the \textit{global memory}, which is accessible by all threads. The next layer is \textit{shared memory}, which is a fast and programmable cache. All the threads in the same thread block have access to the same shared memory. Lastly, the fastest layer is the thread-private register to each thread. To achieve good performance, CUDA programmers must effectively utilize the memory subsystem. For example, when threads in a warp request contiguous global-memory locations, these requests can be aggregated into a single transaction (called \textit{coalesced memory access}); non-coalesced memory access will cause a significant performance slowdown.

\subsection{LZSS}
\label{sec:lzss}
LZSS is a variant of LZ77~\cite{ziv1977universal}, the first algorithm in the LZ compression family. LZSS has the same fundamental idea as other LZ algorithms: search through a sliding window for the longest possible sub-sequence match and encode all identified matches. To clearly explain the LZSS algorithm, we introduce some basic concepts as follows.
\begin{itemize}[noitemsep, topsep=2pt, leftmargin=1.3em]
	\item \textbf{Input stream} is the sequence of bytes to be compressed.
	\item \textbf{Symbol} is the single-/multi-byte unit of the input stream.
	\item \textbf{Look-ahead buffer} is the byte sequence from the coding position to the end of the input stream.
	\item \textbf{Coding position} is the byte position in the input stream currently encoded in the look-ahead buffer.
	\item \textbf{Sliding window} is a buffer (of size $W$), which is the number of bytes from the coding position backward. The window is empty at the beginning, then grows to size $W$ as the input stream is processed, and ``slides'' along with the coding position.
	\item \textbf{Pointer} contains two numbers:
	      the first one is the length of the match, and the second one is the starting offset. The starting offset is the count of bytes from the coding position back to the window, and the length is the number of bytes to read forward from the starting offset.
	\item \textbf{Literal} represents the current byte if there is no match.
	\item \textbf{Flag array}'s each bit indicates whether its corresponding bytes (in compressed data) represent a pointer or a literal.
\end{itemize}

The basic steps of LZSS can be summarized as follows.
\begin{enumerate}[label=\arabic*), noitemsep, topsep=2pt, leftmargin=1.3em]
	\item Set the coding position to the start of the input stream;
	\item Find the longest match started from the coding position; 
	\item If a match is found, output the pointer $P$ and move the coding position and the sliding window $L$ bytes forward, where $L$ denotes the length of the match;
	\item If no match is found, output the first byte in the look-ahead buffer and move the coding position and the sliding window one byte forward;
	\item Use a flag array to record whether a match is a found; 
	\item If the look-ahead buffer is not empty, return to Step 2).
\end{enumerate}

\begin{figure}[t]
	\centering
	\includegraphics[width=\linewidth]{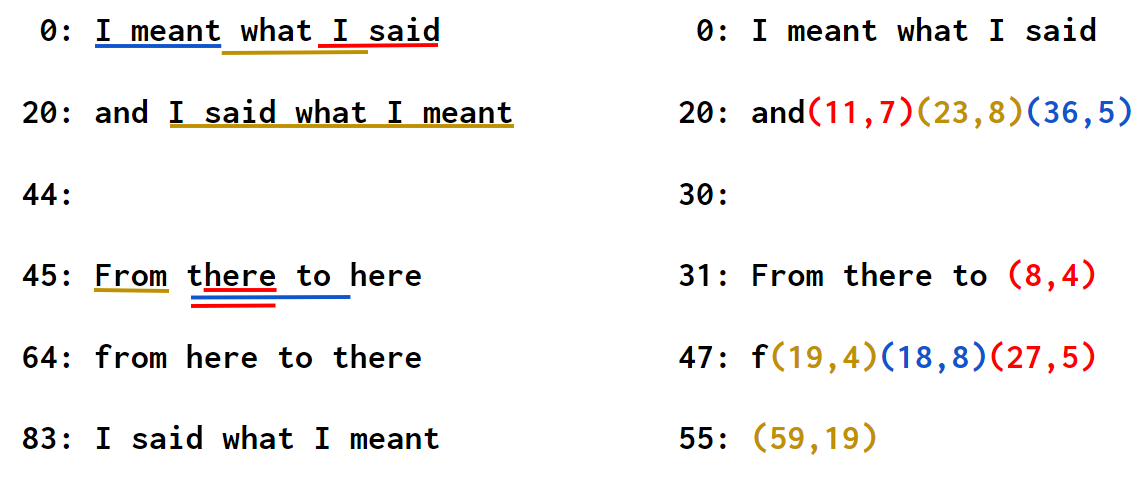}
	\caption{An example of LZSS algorithm. The left is original data, and the right is compressed data. Two numbers in brackets denote length and offset.}
	\label{fig:lzss}
\end{figure}

\FIG~\ref{fig:lzss} illustrates a simple example to demonstrate how LZSS works.
Specifically, the original 102 bytes are compressed to 56, bytes including a flag array. A pair of numbers in brackets represents a pointer, where the first is the offset and the second is the length. If $W$ (also the maximum match length) is less than 256, both the offset and the length can be represented in one byte.
This example demonstrates why LZSS is well suited for compressing data with many repeated patterns.
However, it also indicates LZSS's strong sequential execution characteristics, since the current match must start from the coding position determined by the last match. Due to this strong dependency, LZSS cannot fully leverage the massive parallelism of GPU for high performance.

\subsection{GPU LZ Compression}
CULZSS~\cite{ozsoy2011culzss, culzss} is a state-of-the-art GPU implementation of the LZSS algorithm. It first partitions the input data into multiple chunks to increase the parallelism and then launches a matching kernel on the GPU and an encoding kernel on the CPU. Specifically, the matching kernel lets each GPU thread find the longest match for each byte of the input stream and stores all matches in the global memory,
After that, all matches will be copied from the GPU to the CPU.
Finally, the CPU encoding kernel will sequentially process these matches like the original LZSS.
Note that not all matches will be used in the encoding: if one match covers the following matches, these overlapped matches will be skipped.
Furthermore, Ozsoy \textit{et al.}~\cite{ozsoy2012pipelined} improved CULZSS by overlapping the GPU and CPU computations to increase the performance.
In addition, CULZSS-Bit~\cite{ozsoy2014culzss} adapted CULZSS to handle bit-wise symbols.

We also note that the nvCOMP library~\cite{nvcomp} developed by NVIDIA provides a series of lossless compressors on the GPU, including LZ4. LZ4 is a fast LZ compression implementation, especially featuring fast decoding. Unlike LZSS, LZ4 does not use a flag array to indicate the match pointer; instead, it uses a fixed format with a token (including literal and match length) to save each match.
\begin{figure}[ht]
	\centering
	\includegraphics[width=\linewidth]{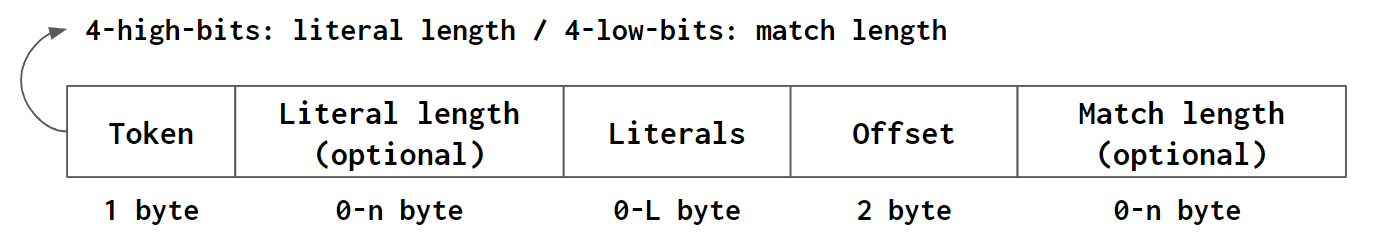}
	\caption{LZ4 format.}
	\label{fig:lz4}
\end{figure}
However, compared to LZSS's flag array, the fixed-length token may result in a lower compression ratio, especially when the length of continuous literals is relatively short. More importantly, we cannot modify the proprietary nvCOMP to accommodate multi-byte data. Thus, in this work, we focus on \culzss{} as our major comparison baseline.

\subsection{Issues of \culzss}
\label{sec:issues}
\culzss{} faces several critical issues:
\Circled{1} \textbf{No support of multi-byte data}: \culzss{} treats the input as a sequence of single bytes regardless of the data type.
This can lead to lower compression ratios because patterns are in multi-byte units and/or lower compression throughput due to the higher computational complexity of searching for matches in units of single bytes.
\Circled{2} \textbf{Fixed data chunk size}: Data chunk size highly impacts the compression ratio and throughput, but it is a fixed value in \culzss. Thus, it is challenging to adapt \culzss{} to different GPU architectures with different shared memory sizes.
\Circled{3} \textbf{Fixed sliding window size:} \culzss{} uses a fixed sliding window size, which prevents a potential tradeoff between compression ratio and throughput.
\Circled{4} \textbf{Under-utilization of shared memory:} \culzss{} underutilizes the GPU shared memory, resulting in multiple buffer updates in shared memory.
\Circled{5} \textbf{CPU encoding:} \culzss{} copies matches (twice the size of the input stream) from the GPU to the CPU and performs the encoding, which causes a significant performance drop due to data copies and slow sequential CPU encoding.

\section{Our Proposed Design}
\label{sec:design}

In this section, we first overview our \thiswork. Then, we describe our proposed algorithm-level optimizations to solve the above issues. Finally, we present the implementation details of our GPU kernels.

\subsection{Overview of \thiswork{}}
\label{sec:matching-kernel}

\begin{figure}[ht]
	\centering
	\includegraphics[width=\linewidth]{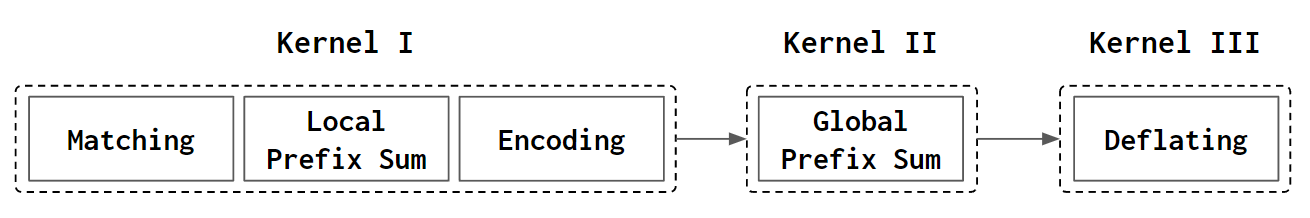}
	\caption{Workflow of our proposed \thiswork{}.}
	\label{fig:gpulz}
\end{figure}

The goal of our design is to fully utilize the GPU resources for high compression throughput and maintain as high in compression ratio as the original/sequential LZSS algorithm.
We illustrate our proposed \thiswork{} in Figure~\ref{fig:gpulz}.
Specifically, \thiswork{} consists of five steps: matching, local (block-level) prefix sum, encoding, global (grid-level) prefix sum, and deflating.
We propose three kernels for these steps. Specifically, Kernel I is for the matching step, the local prefix sum, and encoding to generate the compressed symbols for each data chunk (\S\ref{sec:kernelI}); Kernel II is for the global prefix sum to calculate the memory address (or write offset) of each compressed data chunk in the global memory (\S\ref{sec:kernelII}); and Kernel III is for deflating empty bytes based on the calculated offsets and writing the compressed data (\S\ref{sec:kernelIII}).

Note that we use three separate kernels because a grid-level synchronization across thread blocks is needed to calculate the global memory addresses. In comparison, there is an implicit synchronization between two kernels.
Moreover, although we cannot use a single kernel to handle all computations in the shared memory due to hardware constraints (\S\ref{sec:two_way}), we propose an optimization (i.e., two-pass prefix sum with kernel fusion) that minimizes data movement between the shared and global memories and reduces the global memory footprint (\S\ref{sec:two_way}).

For the matching step, similar to \culzss, we also find the longest match in the sliding window for each symbol in the input stream, even though some matches will not be used in the encoding process.
However, as discussed in \S\ref{sec:issues}, \culzss's matching step does not consider multi-byte symbol,
which significantly degrades the compression ratio and throughput.
To solve this issue, we propose a multi-byte matching approach, which can reduce the computational complexity and find longer matches (\S\ref{sec:multi_byte}).
For the encoding step, as discussed in \S\ref{sec:issues}, \culzss{} must copy matches from the GPU to the CPU and perform the CPU encoding sequentially, leading to low throughput.
This makes \culzss{} impractical for use cases where data generated on the GPU needs to be compressed. Thus, we propose a new compression workflow, including encoding and deflating (see \S\ref{sec:two_way} and \S\ref{sec:enc}, respectively) to get rid of the handling from the CPU side completely.

\subsection{Algorithm-level Optimizations}

Next, we describe our four algorithm-level optimizations in detail.

\subsubsection{Exploring Optimal Workflow}\label{sec:enc}

\begin{figure}[t]
	\centering
	\includegraphics[width=\linewidth]{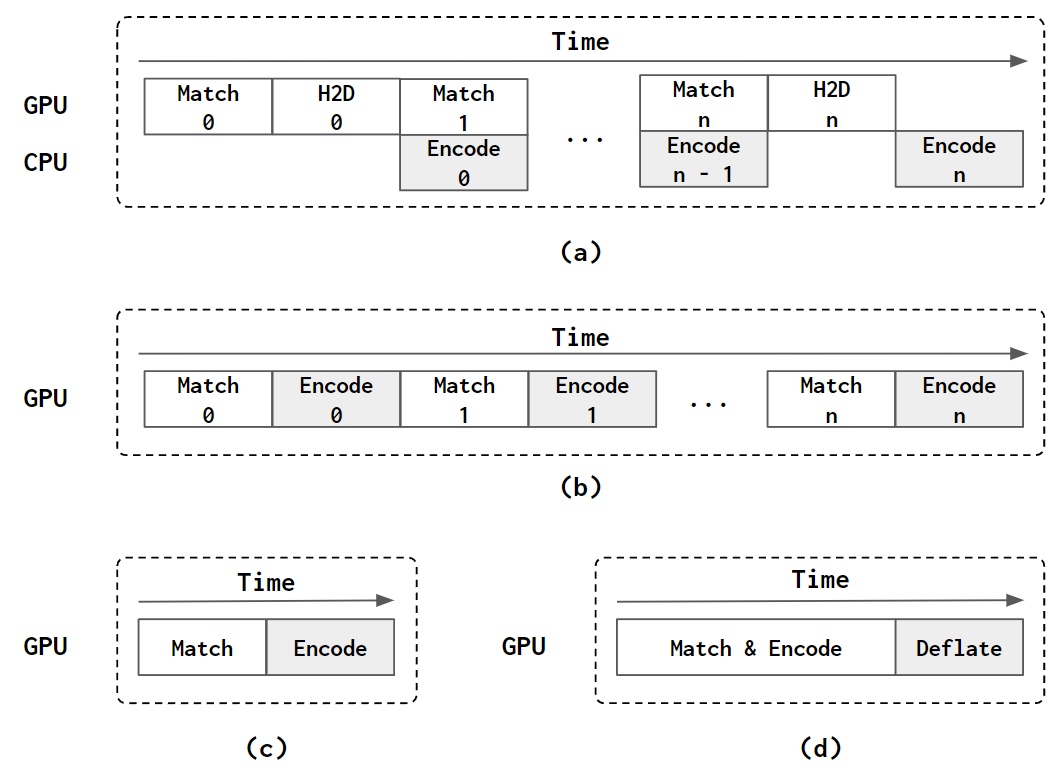}
	\caption{Three workflows of GPU LZSS.}
	\label{fig:encode-diff}
\end{figure}

First, we explore the optimal workflow of LZSS compression on the GPU.
\culzss{} uses the CPU to encode/compress the matches found by the GPU, as shown in Figure~\ref{fig:encode-diff} (a). The CPU encoding kernel and the GPU matching kernel are executed asynchronously to maximize overlapping. However, this workflow makes the encoding process difficult to parallelize, and the GPU-to-CPU data movement is time-consuming. To solve this issue, we propose to perform the encoding on the GPU.
One straightforward solution is to replace the CPU encoding directly with a GPU kernel without changing the workflow, as illustrated in Figure~\ref{fig:encode-diff} (b). However, this workflow is still not optimal because every encoding kernel depends on the last matching kernel. This dependency causes only one kernel to be executed at one time (i.e., sequential execution), which brings the GPU resources starvation problem, especially for small data chunks. Moreover, the multiple kernel launches further increase the time overhead.

To address this issue, we propose our second workflow, as illustrated in Figure~\ref{fig:encode-diff} (c). In this design, we perform the matching and encoding steps in two separate GPU kernels. While the dependency between these two kernels still exists, the size of the data processed is changed from one data chunk to the entire input stream, which solves the starvation problem. However, this design requires a large global memory space to store the intermediate data (i.e., high memory footprint) and causes a large amount of data to be moved between global memory and shared memory.
To this end, we propose our third workflow that performs the matching and encoding steps in the same kernel, as shown in \FIG~\ref{fig:encode-diff} (d). One major issue with fusing the matching and encoding kernels is that the output includes empty bytes. Thus, we need to add another kernel to eliminate these empty bytes (called ``deflating kernel''). This workflow is non-trivial because it is only feasible with our proposed two-pass prefix sum (detailed in \S\ref{sec:two_way}).

\subsubsection{Two-pass Prefix Sum with Kernel Fusion}
\label{sec:two_way}

\begin{figure}[t]
	\centering
	\includegraphics[width=\linewidth]{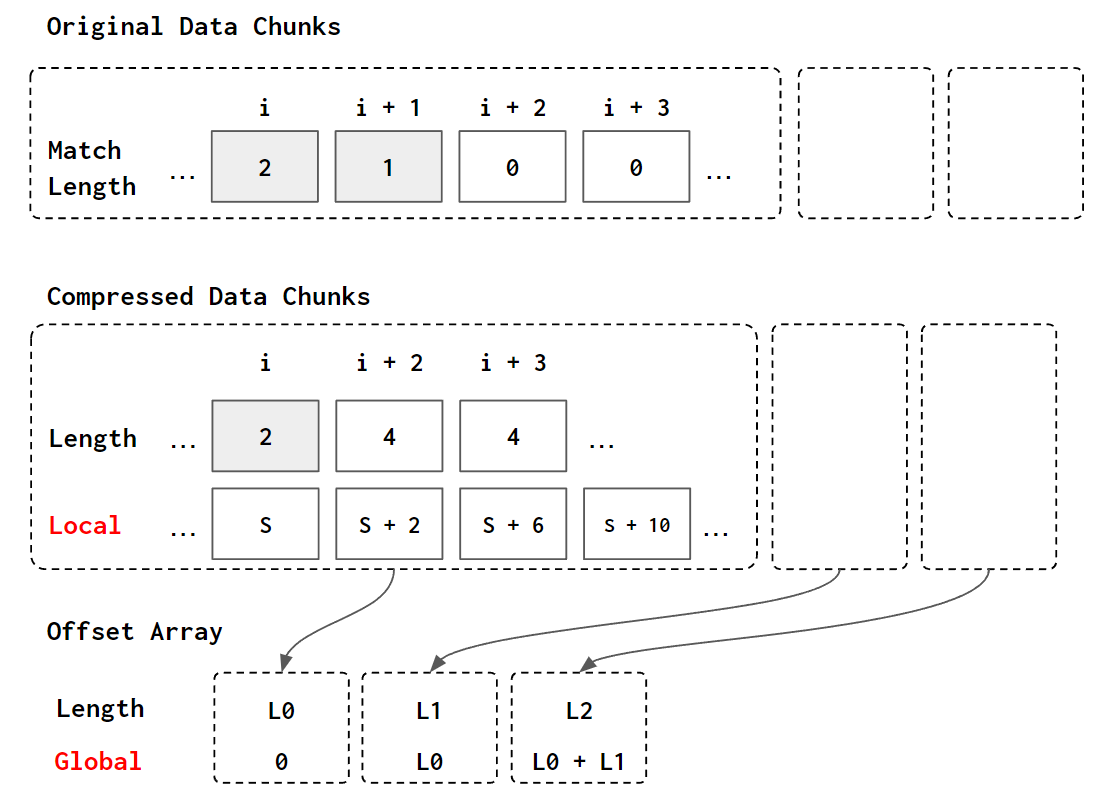}
	\caption{Our proposed two-pass prefix sum.}
	\label{fig:two_way_prefix}
\end{figure}

Fine-grained parallelization of LZSS on GPU is more challenging than coarse-grained parallelization on CPU because GPU thread blocks do not communicate with each other while the kernel function is running, causing the memory offset of each compressed data chunk to be unknown.
To solve this issue, we first locally calculate the size of each symbol after compression (could be either a pointer or a literal), then globally synchronize these sizes, and finally calculate the global offset for each symbol based on a prefix sum.
One approach to achieve global synchronization in a CUDA kernel is to use ``cooperative groups''~\cite{coopgroups}.
However, the cooperative groups API has a limited number of threads (e.g., 1,280 threads on A4000), smaller than we need, which is typically 5,000 threads.
Thus, we need to divide the kernel into two with an implicit device-level synchronization involved.
However, this design requires moving compressed data back to the global buffer, incurring multiple data movements between shared memory and global memory.

To solve this issue, we propose an optimization called two-pass prefix sum. It includes both a local prefix sum and a global prefix sum. The design is shown in Figure~\ref{fig:two_way_prefix}.
Specifically, the local prefix sum calculates the offset for each compressed symbol within each data chunk/thread block. Here we adopt an optimized two-sweep prefix-sum algorithm that fits GPU well~\cite{blelloch1990prefix}. It includes up-sweep and down-sweep processes, detailed in \S\ref{sec:kernelI}.

After we get the compressed size of each data chunk from the local prefix sum, we can calculate the offset of each compressed chunk through a global prefix sum across data chunks.
Compared with the single-pass prefix sum, our proposed two-pass prefix sum only needs to store the size of each compressed data chunk instead of the size of each compressed symbol, which significantly reduces the amount of data written to and read from the global memory (e.g., by at least $C$ times, where $C$ is the number of symbols per data chunk).
Moreover, our two-pass prefix sum can also reduce space complexity and the global memory footprint.
Note that to avoid moving the match result back and forth between the shared and global memories for the local prefix sum, we propose to fuse the local prefix-sum computation into the matching kernel. Thus, we can perform the local prefix-sum computation directly on the matching result stored in the shared memory. This can also reduce the global memory footprint.

After the local prefix sum, each GPU thread encodes symbols based on the calculated local offsets and the found matches, similar to the CPU sequential encoding in LZSS, as mentioned in \S\ref{sec:lzss}.
Note that since this encoding is performed at the thread-block level, no grid-level synchronization is needed. As a result, the encoding can be further fused with the matching and local prefix sum steps to form Kernel I.
It is also worth noting that compared with \culzss, our encoding enables massive GPU threads, which maximizes the parallelism and encoding throughput.

\subsubsection{Multi-byte Matching Approach}
\label{sec:multi_byte}

\culzss{} only performs the matching step on a single-byte basis, which leads to a decrease in compression throughput and a potential loss of compression ratio. This is because, for datasets based on multi-byte symbols, single-byte matching would lose the characteristics of a specific data structure. To this end, we propose a novel multi-byte matching approach that finds matches based on symbols instead of bytes. This strategy has two advantages:
\Circled{1} Searching for matches based on symbols is less expensive than searching for matches based on bytes since there are far fewer symbols than bytes, which increases compression throughput.
\Circled{2} It can bring potentially higher compression ratios because each match can contain more bytes.
We use $S$ to denote the symbol length in the following discussion.

However, the potential gain in compression ratio is not guaranteed, especially when the match length is generally short. Therefore, to maximize the chance of increasing the compression ratio with our multi-byte matching approach, we propose to adaptively select the symbol length and increase the sliding window size.
For example, assuming that the input data type is int32, by default, \thiswork{} adopt the 4-byte symbol length and the sliding window size of 128. Our approach is to adapt the symbol length (ranging from 1 to 4) and the sliding window size (ranging from 32 to 255\footnote{Note that we set the maximum $W$ to 255 because we only use one byte to save the sliding window size and reserve 0 for no match found.}) to achieve the best trade-off between the compression ratio and the throughput.

After studying the impacts of symbol length $S$ and sliding window size $W$ on various datasets (detailed in \S\ref{sec:parameters}), we propose a lightweight parameter selection approach. Specifically, assuming the datasets contain multiple fields that are the input to \thiswork{} at one time, we monitor the average compression ratio with the multi-byte matching strategy (default).
\Circled{1} When the average compression ratio is relatively low (for instance, lower than 1.5), we switch back to single-byte matching,
considering that the multi-byte matching is not effective under low compression ratio circumstances (will be illustrated in \S\ref{sec:parameters}).
This is because the multi-byte matching results in a smaller number of repeated patterns and ignores byte-level repeated patterns.
\Circled{2} When the average compression ratio is high, we keep using multi-byte matching, considering that multi-byte matching has a significant improvement in compression ratio over single-byte matching.
On the other hand, for $W$, we enlarge it to $S$ times when we use the multi-byte matching strategy since the multi-byte matching can bring a speedup of about $S$ times over the single-byte matching, which will offset the higher time complexity brought by a larger sliding window size.

In addition, we provide another option that allows users to set different sliding window sizes, e.g., 32/64/128/255 as level 1-4. A higher level will bring a higher compression ratio but lower throughput. The user can decide the level based on their needs. For example, if compression throughput is a priority, users should select level 1; if the compression ratio is a priority, users should select level 4.

\subsection{Details of \thiswork{} and Its Implementation}
Finally, we introduce our architectural performance optimizations with some implementation details.
We describe these details following our compression workflow. We first introduce the data partition and then describe the kernel details.

\subsubsection{Data Partition}
\label{sec:preprocessing}
\begin{figure}[t]
	\centering
	\includegraphics[width=\linewidth]{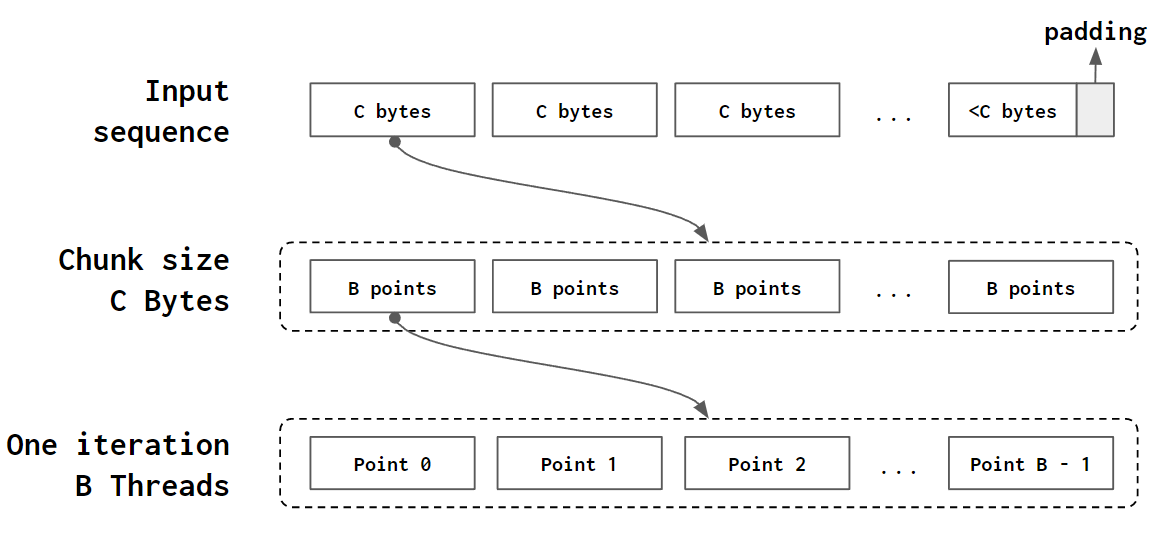}
	\caption{Data partition strategy}
	\label{fig:data_partition}
\end{figure}

First, we divide the input data into multiple blocks and then divide each block into multiple chunks. We launch a GPU kernel for each data block and map each data chunk to one GPU thread block in the kernel. Our data partition strategy is illustrated in Figure~\ref{fig:data_partition}.
The reasons for performing a two-level data partition are as follows:
\Circled{1} \textit{Data block}: GPUs have limited memory capacity (e.g., 16 GBs for an NVIDIA A4000 GPU), so partitioning data into blocks allows the GPU to process datasets that are larger than its memory space.
\Circled{2} \textit{Data chunk}: As aforementioned, the encoding step requires iterating over the found matches and encoding the matches that are not covered by previous matches, introducing data dependencies and sequential execution. Thus, data partitioning can enforce matches not across different chunks, increasing the encoding parallelism.

\paragraph{Data block size:} \culzss{} divides the input data into many small blocks (e.g., 1 MB) such that it overlaps CPU encoding with GPU matching as much as possible. However, this very small block significantly limits the GPU resource utilization and overall computational efficiency.
In contrast, since our design does not involve the CPU for encoding, we can use a relatively large block size to increase GPU efficiency.
Therefore, we choose the block size of 30\% of the global memory (e.g., 12 of 40 GB for NVIDIA A100 GPU).

\paragraph{Data chunk size:} On the one hand, our data partition results in lower compression ratios since we ensure matches cannot span chunks; in other words, larger chunks have higher compression ratios. On the other hand, the chunk size needs to satisfy the GPU hardware constraint as each chunk is stored temporarily in the GPU shared memory. Note that the shared memory is part of the GPU's L1 cache, so the more shared memory is used the less L1 cache is left. The trade-off between shared memory and L1 cache is discussed in detail in \S\ref{sec:parameters}.
In addition, each thread in a thread block processes multiple symbols in a data chunk. We use $C$ to denote the data chunk size in the following discussion.

\subsubsection{Kernel I}
\label{sec:kernelI}
Next, we describe Kernel I and our optimization in warp divergence. We also present its pseudocode in Listing~\ref{alg:matching-kernel}.

\begin{lstlisting}[
    language=c++, 
    morekeywords={__shared__, uint8_t, int, input, output},
    caption={Proposed Kernel I},
    label={alg:matching-kernel}
    ]
input: original data
output: compressed data, flag array, offset arrays

// find matches
for iteration in chunkSize / blockDim.x:
    tid = threadIdx.x + iteration * windowSize
    while windowPointer < bufferStart && bufferPointer < blockSize
        if buffer[bufferPointer] == buffer[windowPointer]:
            if offset[bufferPointer] == 0:
                offset = bufferPointer - windowPointer
            len++
            bufferPointer++
        else:
            if len > maxLen:
                maxLen = len
                maxOffset = offset
            len = 0
            offset = 0
            bufferPointer = bufferStart
        windowPointer++
    writeToSharedMem(lengthArr, offsetArr)
    initializeToZero(prefixArr)
    
// synchronize threads
__syncthreads()

// find encoding information
__shared__ uint8_t byteFlagArr[(dataChunkSize / 8)]

if threadidx.x == 0:
    while encodeIndex < blockSize:
      if lengthBuffer[encodeIndex] < minEncodeLength:
        prefixBuffer[encodeIndex] = sizeof(dataType)
        encodeIndex++
      else:
        prefixBuffer[encodeIndex] = 2
        encodeIndex += lengthBuffer[encodeIndex]
    generateFlagArr(byteFlagArr)

// local prefix sum
localPrefixUpSweep(prefixArr)
saveTheCompressedChunkSize()
localPrefixDownSweep(prefixArr)

// compress data
for iteration in chunkSize / blockDim.x:
    encodeBasedOnlocalPrefix()

// copy flag array back to global mem
writeBackToGlobal(byteFlagArr)
\end{lstlisting}

At the beginning of Kernel I, we load the input stream into the shared memory. Different from \culzss{}, which uses two arrays in the shared memory to store the input stream and the sliding window, we integrate them into one array and use pointers to indicate the start positions of the input stream and sliding window. This design saves the context switch time for updating arrays and fully utilizes the shared memory to accelerate the matching step.

Then, we find matches for every symbol in the designated data chunk.
In the sequential LZSS, the matching process is highly time-consuming. In the best case, it takes O(n) (when no match is found), while in the worst case, it takes O($n^{2}$) time complexity (when the sliding window and the look-ahead buffer contain the same symbol).
This unstable time complexity would cause a severe divergence among different GPU threads. To solve this issue, we use an optimized matching method
with a stable time complexity to reduce the divergence among threads. Our method is described as follows.
\begin{enumerate}[label=\arabic*), noitemsep, topsep=2pt, leftmargin=1.3em]
	\item We use a search pointer to the start of the window and a position pointer to the coding position (Lines 12--25).
	\item We move the search pointer until it reaches the same value as the position pointer points to, and then move both pointers until they point to different values (Lines 13-17).
	\item We record the length and offset of the current match only if it is longer than the previously recorded match (Lines 18--24).
	\item We let the position pointer point to the coding position again, advance the search pointer to the next location, and repeat step 2) until the sliding window is iterated all over or the look-ahead buffer is empty (Lines 12, 22--24).
\end{enumerate}
Although this method cannot guarantee an optimal output (i.e., always finding the longest matches), it gives a sufficient result (will be shown in the evaluation) with a very stable time complexity of O($n$). Moreover, unlike the traditional LZSS, the maximum encoding length in our method does not exceed the offset. Both aspects can minimize the possibility of warp divergence.

After that, we encode the matches (Lines 35-43). Specifically, we use one thread per thread block to calculate the compressed size and produce the flag array (due to LZSS's sequential nature).
Specifically, if a match is found for the current symbol, it takes two bytes (i.e., one for length and one for offset) to encode the match and appends one bit ``1'' to the flag array to denote ``match'' and then skips the symbols that the match covers (Lines 40-42). If no match is found, it saves the original symbol (i.e., the number of bytes for the input data type) and appends one bit ``0'' to the flag array to denote ``no match''
(Lines 37-39).
Note that before the local prefix sum to calculate the memory offsets, we allocate an array in the shared memory to save the size of each compressed symbol (Line 7) and initialize the array to all zeros (Line 27). Thus, we can skip the symbols covered by a match to further improve performance.

\begin{figure}
	\subfloat[Up-sweep]{%
		\includegraphics[clip,width=\columnwidth]{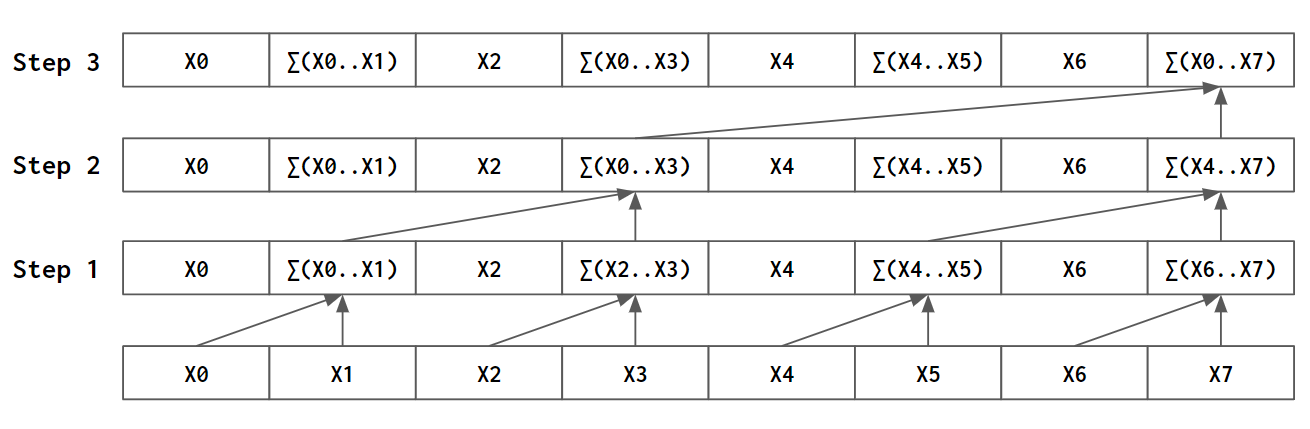}%
	}
	\quad
	\subfloat[Down-sweep]{%
		\includegraphics[clip,width=\columnwidth]{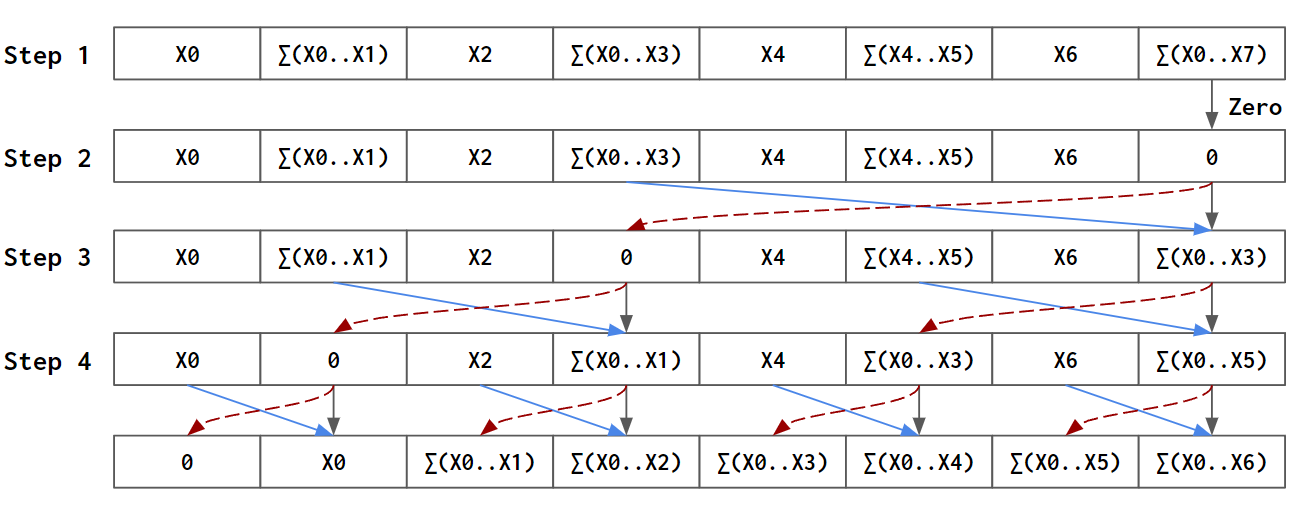}%
	}
	\vspace{2mm}
	\caption{Two-sweep algorithm}
	\label{fig:two-sweep}
\end{figure}

Finally, we use the local prefix sum (described in \S\ref{sec:two_way}) to calculate the memory offset of each symbol within its data chunk and write the compressed data chunks and their sizes to the global memory for deflating.
We use a two-sweep method~\cite{blelloch1990prefix} to implement our local prefix sum, as illustrated in Figure~\ref{fig:two-sweep}. Specifically, in the up-sweep phase, we calculate the summation of two elements with a distance of $2^{\text{step} - 1}$ (Figure~\ref{fig:two-sweep}a). Then, we can get the summation of all elements stored in the last position. After that, we save this back to the global memory for the following global prefix sum and reinitialize it with 0. In the down-seep phase, we copy each element to a new position and calculate the summation of two elements (Figure~\ref{fig:two-sweep}b). Finally, we get the prefix sum of the whole array.

\subsubsection{Kernel II}
\label{sec:kernelII}
Next, we present Kernel II and show its pseudocode in Listing~\ref{alg:prefix}. Specifically, we perform the global prefix sum (discussed in \S\ref{sec:two_way}) in Kernel II. Since prefix sum has been implemented and highly optimized in the CUB library~\cite{cubPrefixSum}, we directly call it to achieve high performance. Note that Kernel II launches the CUB's prefix-sum kernel twice,
because we not only need to calculate the offset of the compressed data (Line 5) but also need to calculate the offset of the flag array (Line 8).

\begin{lstlisting}[
    language=c++,  
    morekeywords={__shared__, uint8_t, int, input, output},
    caption={Proposed Kernel II},
    label={alg:prefix}
    ]
input: compressed size, flag array size 
output: compressed offset, flag array offset

// calcualte the offset for compressed size
cub::DeviceScan::ExclusiveSum(compressedSize, prefix)

// calcualte the offset for flagarray
cub::DeviceScan::ExclusiveSum(flagSize, flagPrefix)
\end{lstlisting}

\subsubsection{Kernel III}
\label{sec:kernelIII}
Finally, we implement the deflating process in Kernel III. The pseudocode of Kernel III is presented in Listing~\ref{alg:encoding-kernel}.
We use the same granularity as the matching step to fully utilize the parallelism of the GPU.
Specifically, we calculate the sizes of the flag array and the compressed data chunk (Lines 5-6) and then write the flag array (Lines 9-11) and the compressed data (Lines 14-16) based on the memory offsets.

\begin{lstlisting}[
    language=c++,  
    morekeywords={__shared__, uint8_t, int, input, output},
    caption={Proposed Kernel III},
    label={alg:encoding-kernel}
    ]
input: compressedData, cOffset, flagArray, fOffset
output: compressedOut, flagArrayOut

int tid = threadIdx.x
flagArrSize = fOffset[Index + 1] - fOffset[Index]
compressedSize = cOffset[Index + 1] - cOffset[Index]

// write back flag array
while tid < flagArrSize:
    write(flagArray, fOffset[index], tid)
    tid += blockDim.x
    
// write back compressed data
while tid < compressedSize:
    write(compressedData, cOffset[index], tid)
    tid += blockDim.x
\end{lstlisting}

\section{Performance Evaluation}\label{sec:evaluation}

In this section, we present our evaluation of {\thiswork} on six representative multi-byte datasets and its comparison with state-of-the-art LZ GPU solutions, i.e., \culzss{} and nvCOMP's LZ4.

\subsection{Experimental Setup}

\paragraph{Platforms.} We use two platforms in our evaluation:
\Circled{1} One node from the Big Red 200 supercomputer~\cite{br200}, equipped with two 64-core AMD EPYC 7742 CPUs @2.25GHz and four NVIDIA Ampere A100 GPUs (108 SMs, 40GB), running CentOS 7.4 and CUDA 11.4.120.
\Circled{2} An in-house workstation equipped with one 24-core
Intel Xeon W-2265 CPU @3.50GHz
and two NVIDIA GTX A4000 GPUs (40 SMs, 16 GB), running Ubuntu 20.04.5 and CUDA 11.7.99.

\paragraph{Datasets.} We conduct our evaluation using six representative multi-byte datasets from two benchmarks, i.e., TPC-H benchmark \cite{tpch} and Scientific Data Reduction Benchmarks (SDRBench)~\cite{sdrbench}.
Specifically, TPC-H benchmark is a suite of business-oriented ad-hoc queries and concurrent data modifications.
It includes one \verb|int32| integer dataset (i.e., tpch-int32) and one utf-8 string dataset (i.e., tpch-string).
SDRBench includes three uint16 datasets (i.e., hurr-quant, hacc-quant, nyx-quant) and one float32 dataset (i.e., rtm).
Note that the three uint16 datasets are intermediate data \footnote{The intermediate data is the quantization code generated by \cusz{}~\cite{cusz} under the relative error bound of 1e-3, and is stored in uint16 format.} generated from three real-world HPC simulation datasets, i.e., HACC (cosmology particle simulation)~\cite{hacc}, Hurricane (ISABEL weather simulation)~\cite{hurr}, and Nyx (cosmology simulation)~\cite{nyx}, which have been widely used in previous studies on scientific data compression~\cite{wang2019compression, tian2020cusz, lu2018understanding, tian2021revisiting, cody2022optimizing, tian2021optimizing, liu2021high}; the \verb|float32| rtm dataset is from a seismic imaging application for petroleum exploration~\cite{kayum2020geodrive,jin2022improving}.

\paragraph{Baselines.} We compare \thiswork with two baselines:
\Circled{1} \textit{\culzss{}}: \culzss{} is the state-of-the-art GPU implementation (open-source)~\cite{culzss} of LZSS, but it uses the GPU to find matches and the CPU to encode matches. \Circled{2} \textit{nvCOMP's LZ4}: LZ4 is similar to LZSS but uses a particular data format to achieve portability. We use the state-of-the-art GPU implementation (closed-source) of LZ4 from nvCOMP~\cite{nvcomp}. We use the latest nvCOMP 2.6.0.

\paragraph{Evaluation metrics.}
We focus on evaluating and analyzing GPU-based LZ compressors on two main metrics.
\Circled{1} \textit{Compression ratio} is one of the most commonly used metrics in compression research. It can be calculated as the ratio of the original data size and reconstructed data size. Higher compression ratios mean denser information aggregation against the original data and faster data transfer.
\Circled{2} \textit{Compression throughput} is the primary consideration when using a GPU-based lossy compressor instead of a CPU-based one. It can be calculated as the ratio of original data size to compression/decompression time. Higher throughput means faster compression and more significant benefits of using compression.

\subsection{Impacts of Parameters $C$, $W$, and $S$}
\label{sec:parameters}

\begin{table*}[!htp]
	\centering
	\caption{Compression ratio of \thiswork{}. Note that some fields are noted as ``n/a'' due to out of the limited shared memory.}\label{tab:compression_ratio}
	\vspace{2mm}
	\footnotesize
	\renewcommand{\arraystretch}{1.1}
\taburulecolor{lightgray}
\begin{tabu}{ @{} lr | *{3}{>{\color{black}} r} | *{3}{r} | *{3}{>{\color{black}} r} | *{3}{r} | }
            &
            &
    \multicolumn{3}{>{\color{black}}c|}{\bfseries chunk size: 2048}
            &
    \multicolumn{3}{c|}{\bfseries chunk size: 4096}
            &
    \multicolumn{3}{>{\color{black}}c|}{\bfseries chunk size: 8192}
            &
    \multicolumn{3}{c|}{\bfseries chunk size: 16384}                                                                                   \\
    \multicolumn{2}{@{}r|}{window size $\downarrow$}
            & 1 byte & 2 bytes & 4 bytes & 1 byte & 2 bytes & 4 bytes & 1 byte & 2 bytes & 4 bytes & 1 byte & 2 bytes & 4 bytes        \\
    \hline
    hurr    & 32     & 3.14    & 3.77    & 3.58   & 3.18    & 3.84    & 3.66   & n/a     & 3.88    & 3.70   & n/a     & n/a     & 3.72 \\
    quant   & 64     & 3.79    & 4.39    & 4.05   & 3.86    & 4.50    & 4.18   & n/a     & 4.56    & 4.25   & n/a     & n/a     & 4.28 \\
            & 128    & 4.39    & 4.91    & 4.44   & 4.51    & 5.09    & 4.64   & n/a     & 5.18    & 4.75   & n/a     & n/a     & 4.81 \\
            & 255    & 4.89    & 5.32    & 4.78   & 5.07    & 5.59    & 5.15   & n/a     & 5.73    & 5.36   & n/a     & n/a     & 5.47 \\
    \hline
    hacc    & 32     & 1.55    & 1.67    & 1.59   & 1.55    & 1.68    & 1.60   & n/a     & 1.68    & 1.61   & n/a     & n/a     & 1.61 \\
    quant   & 64     & 1.71    & 1.82    & 1.71   & 1.72    & 1.84    & 1.73   & n/a     & 1.85    & 1.74   & n/a     & n/a     & 1.75 \\
            & 128    & 1.87    & 1.97    & 1.83   & 1.88    & 2.00    & 1.86   & n/a     & 2.02    & 1.88   & n/a     & n/a     & 1.89 \\
            & 255    & 2.01    & 2.12    & 1.92   & 2.03    & 2.18    & 1.99   & n/a     & 2.20    & 2.03   & n/a     & n/a     & 2.05 \\
    \hline
    nyx     & 32     & 3.97    & 5.07    & 4.80   & 4.04    & 5.20    & 4.95   & n/a     & 5.27    & 5.02   & n/a     & n/a     & 5.06 \\

    quant   & 64     & 5.06    & 6.18    & 5.73   & 5.19    & 6.42    & 6.00   & n/a     & 6.54    & 6.14   & n/a     & n/a     & 6.21 \\
            & 128    & 6.14    & 7.19    & 6.52   & 6.36    & 7.57    & 6.99   & n/a     & 7.79    & 7.25   & n/a     & n/a     & 7.38 \\
            & 255    & 7.08    & 8.03    & 7.11   & 7.46    & 8.65    & 7.94   & n/a     & 9.01    & 8.42   & n/a     & n/a     & 8.64 \\
    \hline
    tpch    & 32     & 1.31    & 1.25    & 1.29   & 1.32    & 1.26    & 1.30   & n/a     & 1.26    & 1.30   & n/a     & n/a     & 1.30 \\
    int32   & 64     & 1.37    & 1.30    & 1.34   & 1.38    & 1.31    & 1.35   & n/a     & 1.31    & 1.35   & n/a     & n/a     & 1.36 \\
            & 128    & 1.43    & 1.34    & 1.38   & 1.44    & 1.35    & 1.39   & n/a     & 1.36    & 1.40   & n/a     & n/a     & 1.41 \\
            & 255    & 1.50    & 1.38    & 1.41   & 1.51    & 1.39    & 1.43   & n/a     & 1.40    & 1.44   & n/a     & n/a     & 1.45 \\
    \hline
    tpch    & 32     & 1.55    & 1.58    & 1.46   & 1.56    & 1.59    & 1.47   & n/a     & 1.60    & 1.48   & n/a     & n/a     & 1.48 \\
    string  & 64     & 2.02    & 1.96    & 1.72   & 2.04    & 1.99    & 1.76   & n/a     & 2.01    & 1.78   & n/a     & n/a     & 1.79 \\
            & 128    & 2.57    & 2.43    & 2.03   & 2.62    & 2.50    & 2.12   & n/a     & 2.54    & 2.17   & n/a     & n/a     & 2.20 \\
            & 255    & 3.08    & 2.84    & 2.27   & 3.19    & 3.00    & 2.47   & n/a     & 3.09    & 2.58   & n/a     & n/a     & 2.64 \\
    \hline
    rtm     & 32     & 2.45    & 2.72    & 2.88   & 2.47    & 2.75    & 2.91   & n/a     & 2.77    & 2.93   & n/a     & n/a     & 2.94 \\
    float32 & 64     & 2.59    & 2.80    & 2.92   & 2.61    & 2.83    & 2.96   & n/a     & 2.85    & 2.98   & n/a     & n/a     & 2.99 \\
            & 128    & 2.66    & 2.84    & 2.94   & 2.69    & 2.88    & 2.99   & n/a     & 2.89    & 3.01   & n/a     & n/a     & 3.02 \\
            & 255    & 2.69    & 2.85    & 2.97   & 2.72    & 2.90    & 3.02   & n/a     & 2.92    & 3.05   & n/a     & n/a     & 3.07 \\
    \bottomrule
\end{tabu}

\end{table*}

First, we evaluate the impacts of parameters $C$, $W$, and $S$.
We conduct the experiments on both the A100 and A4000 platforms. The compression ratio is shown in Table~\ref{tab:compression_ratio}, and the compression throughput is shown in Table~\ref{tab:compression_tp}.
Specifically, we choose the data chunk sizes (i.e., $C$) of 2048, 4096, 8192, and 16,384. The data chunk size directly decides the shared memory size we utilize in our design. Because the shared memory is part of the L1 cache. As a result, we can observe the impact of the trade-off between shared memory and L1 cache on the overall throughput. Note that some fields in the table are empty because of the limited shared memory.
The sliding window size $W$ will directly decide the time complexity. The longer the sliding window is, the higher the time complexity will be. It will also potentially increase the compression ratio.
Moreover, we introduce multi-byte symbols into the LZSS algorithm to explore the potential compression ratio and throughput gains. To this end, we select three symbol lengths (i.e., $S$): 1, 2, and 4 bytes.

First, we focus on the impact of $C$. As mentioned before, we partition the data into chunks to allow LZSS to execute in parallel. However, due to the independence of each data chunk, the compression ratio would drop slightly because the match does not span the boundaries of data chunks, leading to the limited match length. The evaluation result also proves this, as illustrated in Table~\ref{tab:compression_ratio}. The compression ratio increases as the data chunk size increases in all test cases. The average improvement is 1.02$\times$. However, as the data chunk size increases, the compression throughput decreases in almost all test cases. This proves that a larger L1 cache is better for compression throughput than utilization of shared memory, at least in the range of feasible data chunk sizes of \thiswork{}. With smaller data chunk size, the compression throughput is improved by 1.33$\times$ on average. Note that the compression throughput drops significantly with larger data chunk sizes.
For example, on the 4-byte nyx-quantization dataset, the compression throughput drops from 19.05 to 18.76 when the data chunk size changes from 2048 to 4096. At the same time, it drops from 14.67 to 8.36 when the data chunk size changes from 8192 to 16,384. This is because when the data chunk size is 16,384, the shared memory size is close to the hardware's limit, resulting in a fairly small L1 cache size and further impacting the overall throughput. This phenomenon is more obvious when the data chunk size is bigger. Note that A100 has a higher speedup than A4000 when the block size is large because A100 has larger L1 cache (192 KB/SM) than A4000 (128 KB/SM).

Next, we explore the impact of $W$. On the one hand, as analyzed before, a larger sliding window brings a potentially longer match, increasing the compression ratio. Table~\ref{tab:compression_ratio} shows that the ratio of compression ratio to the sliding window size is near linearly
in almost all datasets.
For example, on the 2-byte tpch-int32 dataset, the compression ratio is 1.26, 1.31, 1.35, and 1.39 when the sliding window size is 32, 64, 128, and 255, respectively. Moreover, the overall compression ratio improvement by extending the sliding window size from 32 to 255 is 1.4$\times$.
On the other hand, a larger sliding window incurs more operations per thread, decreasing the compression throughput.
The average speedup when we change the sliding window size from 255 to 32 is 3.9$\times$. Compared with the relatively small increase in compression ratio, the throughput decreases dramatically as the sliding window size doubles. However, we find \thiswork{} highly stable in throughput across different datasets under the same configuration (i.e., $C$, $W$, and $S$). For example, the throughput of $C=2048$, $W=32$, and $S=2$ is 9.57 GB/s, 8.49 GB/s, 10.14 GB/s, 8.3 GB/s, 7.96 GB/s, and 9 GB/s on A4000 on \{hurr, hacc, nyx\}-quant, tpch-\{int32, string\}, and rtm datasets, respectively.

\begin{table*}[!htp]
	\centering
	\caption{Compression throughput of \thiswork{} on both A100 (blue) and A4000 (gray) GPUs. The red bars show the performance gain when scaling from A4000 to A100.
	}
	\vspace{2mm}
	\label{tab:compression_tp}
	\footnotesize
	\resizebox{\linewidth}{!}{%
		\renewcommand*{\arraystretch}{1.1}%
\taburulecolor{lightgray}
\begin{tabu}{ @{} lr | *{4}{ r@{}r@{}r | } 
        @{}}
                &
                &
        \multicolumn{3}{c| }{\bfseries chunk size: 2048}
                &
        \multicolumn{3}{c| }{\bfseries chunk size: 4096}
                &
        \multicolumn{3}{c| }{\bfseries chunk size: 8192}
                &
        \multicolumn{3}{c@{} |}{\bfseries chunk size: 16384}                                                                                                                                                                                          \\
        \multicolumn{2}{@{}r| }{window size $\downarrow$}
                & 1 byte & 2 bytes         & 4 bytes           & 1 byte            & 2 bytes         & 4 bytes          & 1 byte            & 2 bytes       & 4 bytes          & 1 byte            & \ \ 2 bytes   & 4 bytes                          \\
        \hline
        hurr    & 32     & \TP{8.13}{4.87} & \TP{14.85}{9.57}  & \TP{29.00}{18.07} & \TP{6.87}{3.10} & \TP{14.82}{8.38} & \TP{27.98}{17.44} & \notavailable & \TP{11.26}{4.37} & \TP{26.62}{13.29} & \notavailable & \notavailable & \TP{15.99}{7.55} \\
        quant   & 64     & \TP{4.63}{2.93} & \TP{8.90}{5.55}   & \TP{17.53}{11.18} & \TP{4.47}{2.19} & \TP{8.64}{5.28}  & \TP{17.15}{10.95} & \notavailable & \TP{7.40}{3.11}  & \TP{16.64}{9.22}  & \notavailable & \notavailable & \TP{11.84}{5.62} \\
                & 128    & \TP{2.52}{1.60} & \TP{4.87}{3.08}   & \TP{11.04}{6.73}  & \TP{2.43}{1.33} & \TP{4.72}{2.94}  & \TP{10.08}{6.31}  & \notavailable & \TP{4.30}{1.95}  & \TP{9.45}{5.68}   & \notavailable & \notavailable & \TP{7.40}{3.71}  \\
                & 255    & \TP{1.39}{0.87} & \TP{2.76}{1.75}   & \TP{6.95}{4.37}   & \TP{1.33}{0.79} & \TP{2.60}{1.60}  & \TP{5.70}{3.63}   & \notavailable & \TP{2.34}{1.13}  & \TP{5.30}{3.26}   & \notavailable & \notavailable & \TP{4.33}{2.25}  \\
        \hline
        hacc    & 32     & \TP{7.35}{4.16} & \TP{13.81}{8.49}  & \TP{28.96}{18.14} & \TP{5.83}{2.63} & \TP{13.13}{7.54} & \TP{27.48}{15.49} & \notavailable & \TP{9.31}{3.42}  & \TP{24.57}{10.58} & \notavailable & \notavailable & \TP{14.50}{6.06} \\
        quant   & 64     & \TP{4.47}{2.75} & \TP{8.21}{5.31}   & \TP{19.19}{11.44} & \TP{4.10}{2.08} & \TP{8.15}{5.10}  & \TP{19.07}{11.42} & \notavailable & \TP{6.57}{2.74}  & \TP{17.36}{8.39}  & \notavailable & \notavailable & \TP{11.12}{5.27} \\
                & 128    & \TP{2.56}{1.69} & \TP{4.76}{3.05}   & \TP{12.36}{6.31}  & \TP{2.59}{1.37} & \TP{4.65}{3.04}  & \TP{11.06}{6.74}  & \notavailable & \TP{4.21}{1.98}  & \TP{11.12}{6.03}  & \notavailable & \notavailable & \TP{7.87}{3.58}  \\
                & 255    & \TP{1.50}{0.95} & \TP{2.68}{1.83}   & \TP{7.40}{4.37}   & \TP{1.50}{0.82} & \TP{2.68}{1.67}  & \TP{6.65}{3.93}   & \notavailable & \TP{2.49}{1.21}  & \TP{6.00}{3.51}   & \notavailable & \notavailable & \TP{4.93}{2.45}  \\
        \hline
        nyx     & 32     & \TP{9.53}{6.04} & \TP{15.65}{10.14} & \TP{30.08}{19.05} & \TP{7.53}{3.95} & \TP{15.75}{9.05} & \TP{30.26}{18.76} & \notavailable & \TP{12.44}{5.62} & \TP{29.18}{14.67} & \notavailable & \notavailable & \TP{18.13}{8.36} \\
        quant   & 64     & \TP{5.69}{3.57} & \TP{9.44}{6.15}   & \TP{19.81}{11.60} & \TP{5.40}{2.83} & \TP{9.30}{6.15}  & \TP{18.00}{11.40} & \notavailable & \TP{8.12}{3.75}  & \TP{17.94}{10.79} & \notavailable & \notavailable & \TP{12.90}{6.33} \\
                & 128    & \TP{3.05}{1.92} & \TP{5.48}{3.57}   & \TP{11.29}{7.10}  & \TP{3.08}{1.66} & \TP{5.91}{3.38}  & \TP{10.31}{6.77}  & \notavailable & \TP{5.01}{2.47}  & \TP{10.16}{6.45}  & \notavailable & \notavailable & \TP{8.71}{4.61}  \\
                & 255    & \TP{1.76}{1.03} & \TP{3.56}{2.07}   & \TP{6.94}{4.86}   & \TP{1.67}{0.91} & \TP{3.17}{1.90}  & \TP{6.60}{4.13}   & \notavailable & \TP{3.07}{1.40}  & \TP{6.30}{3.87}   & \notavailable & \notavailable & \TP{5.29}{2.82}  \\
        \hline
        tpch    & 32     & \TP{7.07}{3.93} & \TP{12.06}{8.30}  & \TP{25.42}{14.87} & \TP{5.21}{2.25} & \TP{11.66}{6.21} & \TP{21.40}{13.49} & \notavailable & \TP{7.67}{3.09}  & \TP{19.39}{9.29}  & \notavailable & \notavailable & \TP{10.45}{5.01} \\
        int32   & 64     & \TP{4.38}{2.68} & \TP{7.87}{5.11}   & \TP{16.32}{10.20} & \TP{3.76}{1.74} & \TP{7.53}{4.52}  & \TP{14.58}{9.78}  & \notavailable & \TP{5.89}{2.38}  & \TP{14.24}{7.05}  & \notavailable & \notavailable & \TP{8.23}{4.19}  \\
                & 128    & \TP{2.41}{1.57} & \TP{4.75}{3.02}   & \TP{10.22}{6.31}  & \TP{2.24}{1.14} & \TP{4.46}{2.76}  & \TP{9.23}{5.56}   & \notavailable & \TP{3.81}{1.68}  & \TP{8.39}{5.01}   & \notavailable & \notavailable & \TP{6.44}{3.11}  \\
                & 255    & \TP{1.32}{0.85} & \TP{2.76}{1.72}   & \TP{6.67}{3.97}   & \TP{1.24}{0.68} & \TP{2.41}{1.57}  & \TP{5.76}{3.45}   & \notavailable & \TP{2.14}{1.02}  & \TP{4.99}{3.08}   & \notavailable & \notavailable & \TP{3.77}{2.00}  \\
        \hline
        tpch    & 32     & \TP{7.08}{4.23} & \TP{12.52}{7.96}  & \TP{22.92}{13.82} & \TP{5.32}{2.46} & \TP{11.39}{6.16} & \TP{21.11}{12.62} & \notavailable & \TP{7.98}{3.51}  & \TP{18.95}{8.42}  & \notavailable & \notavailable & \TP{9.96}{4.83}  \\
        string  & 64     & \TP{4.71}{3.06} & \TP{8.35}{5.21}   & \TP{15.15}{9.50}  & \TP{4.16}{1.97} & \TP{7.58}{5.09}  & \TP{15.56}{9.23}  & \notavailable & \TP{6.29}{2.88}  & \TP{13.98}{6.83}  & \notavailable & \notavailable & \TP{8.22}{4.13}  \\
                & 128    & \TP{2.40}{1.74} & \TP{4.77}{3.28}   & \TP{10.71}{5.97}  & \TP{2.57}{1.31} & \TP{4.71}{3.13}  & \TP{9.43}{5.68}   & \notavailable & \TP{3.98}{1.98}  & \TP{8.15}{4.84}   & \notavailable & \notavailable & \TP{5.74}{3.50}  \\
                & 255    & \TP{1.41}{0.94} & \TP{2.58}{1.83}   & \TP{6.91}{3.74}   & \TP{1.42}{0.79} & \TP{2.62}{1.72}  & \TP{6.06}{3.31}   & \notavailable & \TP{2.31}{1.19}  & \TP{4.74}{3.33}   & \notavailable & \notavailable & \TP{3.68}{2.26}  \\
        \hline
        rtm     & 32     & \TP{7.31}{4.23} & \TP{14.27}{9.00}  & \TP{28.35}{17.63} & \TP{6.05}{2.79} & \TP{13.92}{7.54} & \TP{28.57}{17.17} & \notavailable & \TP{11.22}{3.91} & \TP{26.63}{13.14} & \notavailable & \notavailable & \TP{16.23}{7.44} \\
        float32 & 64     & \TP{4.48}{2.85} & \TP{9.27}{5.52}   & \TP{17.69}{11.15} & \TP{3.92}{2.18} & \TP{8.30}{5.10}  & \TP{17.40}{10.90} & \notavailable & \TP{7.02}{3.04}  & \TP{16.77}{9.08}  & \notavailable & \notavailable & \TP{12.37}{5.52} \\
                & 128    & \TP{2.46}{1.77} & \TP{4.89}{3.38}   & \TP{10.81}{6.84}  & \TP{2.37}{1.39} & \TP{4.71}{3.07}  & \TP{9.99}{6.35}   & \notavailable & \TP{4.17}{1.89}  & \TP{9.59}{5.61}   & \notavailable & \notavailable & \TP{7.52}{3.65}  \\
                & 255    & \TP{1.35}{0.98} & \TP{3.07}{2.03}   & \TP{8.05}{4.82}   & \TP{1.31}{0.82} & \TP{3.06}{1.80}  & \TP{6.08}{3.82}   & \notavailable & \TP{2.55}{1.28}  & \TP{5.78}{3.28}   & \notavailable & \notavailable & \TP{4.47}{2.57}  \\
        \hline
\end{tabu}

	}
	\vspace{-2mm}
\end{table*}

Finally, we discuss the impact of the symbol length $S$. As mentioned, the multi-byte symbol length can introduce a potential compression ratio improvement and increase the compression throughput due to longer matches and fewer symbols to process. Table~\ref{tab:compression_ratio} shows that the compression ratio improvement is not determined as we expected. It has different patterns for different datasets. For example, on the three uint16 quantization-code datasets, the compression ratio reaches the peak at $S=2$, which is the same as the length of uint16. However, on the int32 tpch-int32 dataset, the compression ratio is optimal at $S=1$, which is different from the length of int32. This is because the number of repeated patterns is relatively smaller in the tpch-int32 dataset, as indicated by the low compression ratio. Thus, using a 1-byte symbol (i.e., $S=1$) may detect more byte-level repeated patterns and achieve a higher compression ratio than using a 4-byte symbol.
On the utf-8 tpch-string dataset and the float32 rtm dataset, the best compression ratio is achieved at $S=1$ and $S=4$, respectively, which is the same as the unit length of their data types.

Regarding throughput, the impact of the symbol length is more obvious; that is, longer symbol results in higher throughput. The average throughput improvement is 4.5$\times$ when we change $S$ from 1 to 4. Combined with the above observation regarding compression ratio, we find that $S=2$ has both a higher compression ratio and throughput than $S=1$ in some cases. For example, on the hurr-quantization dataset with any $W$ and $C$, $S=2$ can always lead to a better compression ratio and throughput than $S=1$. Note that this observation can be generalized to all LZ compressors.

\subsection{Evaluation on Compression Ratio}
\label{sec:ratio-comp}
\begin{figure*}[t]
	\centering
	\includegraphics[width=\linewidth]{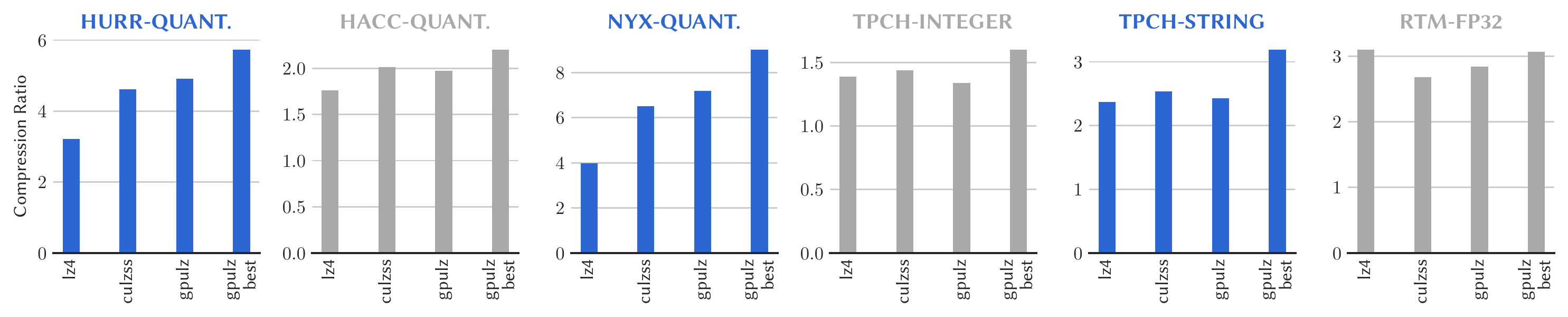}
	\caption{Compression ratio of different GPU compressors.}
	\label{fig:cr}
\end{figure*}

Next, we compare the compression ratio of \thiswork{} with \culzss{} and nvCOMP's LZ4, as shown in Figure~\ref{fig:cr}.
Note that in the figure, we use ``gpulz'' to denote the default configuration ($C=2048$, $S=2$, and $W=128$) and ``gpulz-best'' to denote the best compression ratio from all settings.
The figure shows that compared with \culzss{}, \thiswork{} achieves a similar compression ratio on all datasets because the compression ratio is highly dependent on the sliding window size. In our default configuration, we use $W=128$ as the same as the \culzss{}.
In the best cases (overall configurations), \thiswork{} has an improvement of 1.4$\times$ on compression ratio.
Compared with nvCOMP's LZ4, \thiswork{} achieves an average compression ratio improvement of 1.23$\times$. Specifically, on the hurr-quantization and nyx-quantization datasets, \thiswork{} has the highest compression ratio improvements, which are 1.53$\times$ and 1.8$\times$, respectively. In the best cases (of all configurations), \thiswork{} has an improvement of 1.42$\times$ on compression ratio thanks to our fine-tuned parameters.

\begin{figure*}[t]
	\centering
	\includegraphics[width=\linewidth]{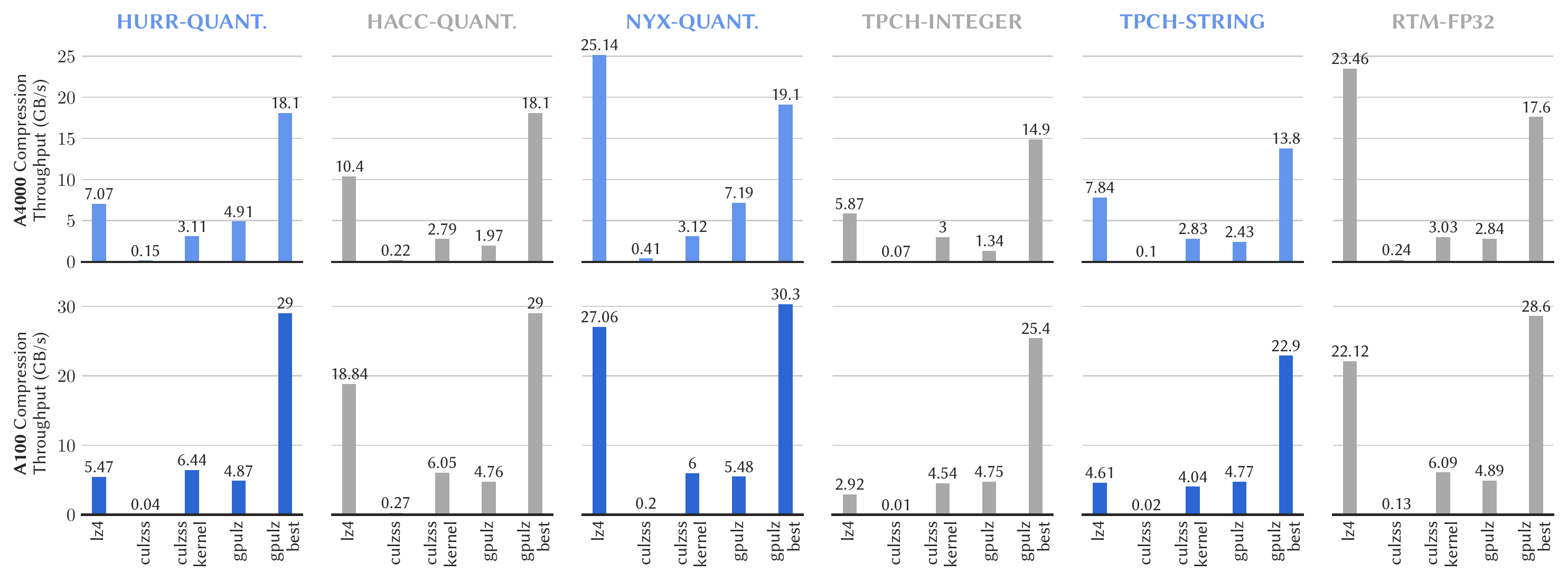}
	\caption{Compression throughput of different GPU compressors on A100 and A4000.}
	\label{fig:tp_comparison}
\end{figure*}

\subsection{Evaluation on Compression Throughput}

Then, we evaluate the performance of \thiswork{}, with its scalability on A100 and A4000 shown in Figure~\ref{fig:tp_comparison}.
Note that ``culzss'' denotes the overall throughput of \culzss{}, including the GPU matching kernel and the CPU encoding process. ``culzss-kernel'' denotes the throughput of the GPU matching kernel. ``gpulz'' denotes our method with the default setting ($C=2048$, $S=2$, and $W=128$). ``gpulz-best'' denotes the best compression throughput from all settings.
We note that the settings for the best case are generally C=2048, W=32, S=4, except for the nyx-quantization and rtm datasets on A100, which achieve the best performance with C=4096, W=32, S=4. This is because A100 has a larger L1 cache (192 KB/SM) than A4000 (128 KB/SM); a larger chunk size (i.e., more shared memory utilization) will not significantly affect performance. Compression throughput is potentially increased by fully utilizing the high-speed shared memory.

Compared with \culzss{}, our method has an average speedup of 22.19$\times$ on all datasets with A4000. When compared with our best case, this speedup increases to 130.01$\times$. The reason is three-fold: \Circled{1} the slow CPU encoding process, \Circled{2} the data movement overhead between CPU and GPU, and \Circled{3} the overhead of multiple times of launching the same kernel. Note that the entire process of \thiswork{} is almost as fast as the matching kernel of \culzss{} both on A100 and A4000, thanks to our optimizations in both algorithm and implementation.
Compared with nvCOMP's LZ4, \thiswork{} has similar compression throughputs on the hurr-quantizaiton, tpch-int32, and tpch-string datasets but slightly slower on the hacc-quantization and nyx-quantization datasets. However, since nvCOMP is not an open-source library, we can only infer the underlying reason, which may be that some field sizes in these datasets are too small.
By comparison, \thiswork{} has more stable performance across all datasets, and our best case achieves higher throughput than nvCOMP with both A100 and A4000 platforms on almost all datasets. For example, the average speedup is 4.32$\times$ on A100.

In addition, we also implement our decompression and evaluate its throughput. Considering decompression is easy to parallel, we do not describe and compare it with other compressors in detail; instead, we only show the average decompression throughput across all datasets.
Specifically, the average decompression throughput of \thiswork{} on all datasets is 16.4 GB/s on A4000 and 29.1 GB/s on A100.
For comparison, nvCOMP's LZ4 has an average decompression throughput of 21.1 GB/s on A4000.

\subsection{Use-case of \thiswork{}}
Finally, we apply \thiswork{} to \cusz{} (a state-of-the-art GPU lossy compressor for scientific data) due to its high performance on the quantization-code datasets to improve the compression ratio. Note that the original \cusz{} only has a Huffman encoding~\cite{huffman1952method}, whereas the improved \cusz{} includes \thiswork{} before the Huffman encoding.
We evaluate the original \cusz{} and the improved \cusz{} on the A100 platform under the relative error bound 1e-2.
Besides Hurricane, NYX, and RTM, we also include one more dataset from SDRBench, i.e., CESM (climate simulation)~\cite{cesm}.

\begin{table}[ht]
	\caption{Comparison of compression ratio and throughput (GB/s) between original \cusz{} and improved \cusz{} (with \thiswork) on A100 platform.}
	\vspace{2mm}
	\resizebox{.32\textwidth}{!}{
		\centering
		\footnotesize
		\begin{tabular}{@{} lrrrrr @{}}
        \toprule
        \bfseries Dataset & \multicolumn{2}{c}{\bfseries \cusz} & \multicolumn{2}{c}{\bfseries \cusz{} w/ \thiswork}
        \\
        \cmidrule(lr){2-3}
        \cmidrule(l){4-5}
                          & \bfseries CR                        & \bfseries THR                                      & \bfseries CR & \bfseries THR
        \\
        \midrule
        CESM              & 22.6                                & 12.0                                               & 43.2         & 2.7           \\
        Hurricane         & 24.3                                & 31.9                                               & 29.1         & 5.9           \\
        Nyx               & 30.1                                & 87.2                                               & 74.8         & 10.4          \\
        RTM               & 28.6                                & 49.2                                               & 249.8        & 7.2           \\
        \bottomrule
\end{tabular}
	}
	\label{tab:end2end}
\end{table}

\TAB~\ref{tab:end2end} shows that the improved \cusz{} obtains an improvement of $1.9 \times \sim 8.7 \times$ in compression ratio with a slightly lower compression throughput.
We note that the improved \cusz{} has higher compression ratio improvements on larger error bounds and higher dimensional datasets (e.g. 3D Hurricane and RTM), since the quantization code generated by \cusz{} in these cases has more spatial redundancy, thus benefiting \thiswork.
Enabling higher compression ratios is critical for many HPC applications using lossy compression (rather than lossless compression). We also note that some CPU lossy compressors with multi-threading support such as SZ~\cite{sz17} and ZFP~\cite{zfp} can also achieve compression throughputs of about 2$\sim$4 GB/s on 32 cores~\cite{dube2022efficient}, but their overall throughput is limited by moving uncompressed data from the GPU to the CPU; in comparison, the time of moving compressed data (with hundreds of compression ratios) with the improved \cusz{} is much lower.

\section{Conclusion and Future Work}\label{sec:conclusion}

In this paper, we propose a series of optimizations for one of the most important lossless compression algorithms LZSS for multi-byte data on GPUs.
Specifically, we develop a new method for multi-byte pattern matching, optimize the prefix-sum operation, and fuse multiple GPU kernels, thereby improving both compression ratio and throughput (due to lower computational time complexity, less data movement, and potentially longer matches).
	{\thiswork} achieves up to 272.1$\times$ speedup and up to 1.4$\times$ higher compression ratio over state-of-the-art solutions.

In the future, we plan to evaluate {\thiswork} on more multi-byte datasets.
We will attempt to develop an analytical model for searching the optimal parameter combination for different datasets.
In addition, we will integrate {\thiswork} into more data-intensive applications running on different parallel and distributed systems.

\section*{Acknowledgment}
This research was supported by the Exascale Computing Project (ECP), Project Number: 17-SC-20-SC, a collaborative effort of two DOE organizations---the Office of Science and the National Nuclear Security Administration, responsible for the planning and preparation of a capable exascale ecosystem, including software, applications, hardware, advanced system engineering and early testbed platforms, to support the nation's exascale computing imperative. The material was supported by the U.S. Department of Energy, Office of Science, Advanced Scientific Computing Research (ASCR), under contract DE-AC02-06CH11357.
This work was also supported by the National Science Foundation under Grants OAC-2003709, OAC-2104023, OAC-2303064, OAC-2247080, and OAC-2312673.
This research was also supported in part by Lilly Endowment, Inc., through its support for the Indiana University Pervasive Technology Institute.

\newpage

\bibliographystyle{plain}
\bibliography{refs}
\end{document}